\documentclass{phbauth}
\usepackage{graphicx}

\begin{document}

\begin{frontmatter}

\title{Aharonov-Bohm effect in circular carbon nanotubes}

\author[address1]{J\"{o}rg Rollb\"{u}hler},
\author[address2,address3]{Arkadi A. Odintsov\thanksref{thank1}},

\address[address1]{Fakult\"{a}t f\"{u}r Physik, Albert-Ludwigs-Universit\"{a}t, 
D-79104 Freiburg, Germany}
\address[address2]{Department of Applied Physics and DIMES,
 Delft University of Technology,  2628 CJ Delft, The Netherlands}
\address[address3]{Nuclear Physics Institute, Moscow State University, 
Moscow 119899 GSP, Russia}

\thanks[thank1]{Corresponding author.
E-mail: odintsov@duttntn.tn. tudelft.nl}

\begin{abstract}
We study the interference of interacting electrons 
in toroidal single-wall carbon nanotubes 
coupled to metallic electrodes by tunnel junctions.
The dc conductance shows resonant features 
as a function of the gate voltage and the magnetic field. 
The conductance pattern is determined by the interaction parameter, 
which in turn can be cross-checked 
against the exponents governing the transport at high temperatures.  
The coordinate dependence of the conductance reflects 
electron correlations in one-dimensional space.
\end{abstract}

\begin{keyword}
Carbon nanotubes; Aharonov-Bohm effect
\end{keyword}
\end{frontmatter}

\bigskip Recent experiments \cite{Liu,Avouris} have revealed ring-shape
structures of single-wall carbon nanotubes (SWNTs). The data by Liu \textit{%
et.al.} \cite{Liu} suggests that the rings consist of ropes of toroidal
SWNTs. Presumably, single toroidal SWNTs have been detected as well. On the
other hand, samples produced by a different technique \cite{Avouris} consist
of ropes of coiled (non-toroidal) SWNTs.

The Coulomb interaction in one-dimensional (1D) SWNTs leads to non-Fermi
liquid electron correlations observed in recent experiments \cite{Bockrath}.
Toroidal SWNTs is an unique system for studying the \emph{interference} of
interacting electrons. In this work we investigate the Aharonov-Bohm effect
in toroidal SWNTs coupled to metallic electrodes by tunnel junctions at the
points $x_{1}=0,$ $x_{2}=x$.

The conductance $G$ of the system can be evaluated in the lowest order in
tunneling, 
\begin{equation}
G=\frac{\pi \hbar G_{1}G_{2}v_{F}^{2}}{96e^{2}}\sum_{s}\left| \hat{G}
_{s}(x,\omega =0)\right| ^{2},  \label{eq:Conductance}
\end{equation}
where $G_{1,2}$ are the conductances of the junctions, $v_{F}$ is the Fermi
velocity in SWNT, and $\hat{G}_{s}$ is the Matsubara's Green's function for
electrons with spin $s$ in SWNT. The latter is determined by the Fermi
operators $\psi _{\alpha sd}$ for right/left ($d=\pm $) moving electrons
near the Fermi points $\alpha K$ ($\alpha =\pm $), 
\begin{equation}
\hat{G}_{s} =-\sum_{\alpha d}e^{i\alpha Kx}\left\langle T_{\tau }\psi
_{\alpha sd}(x,\tau )\psi _{\alpha sd}^{\dagger }(0,0)\right\rangle .
\label{eq:GMats}
\end{equation}
\begin{figure}[tb]
\par
\begin{center}
\leavevmode
\includegraphics[width=0.8\linewidth]{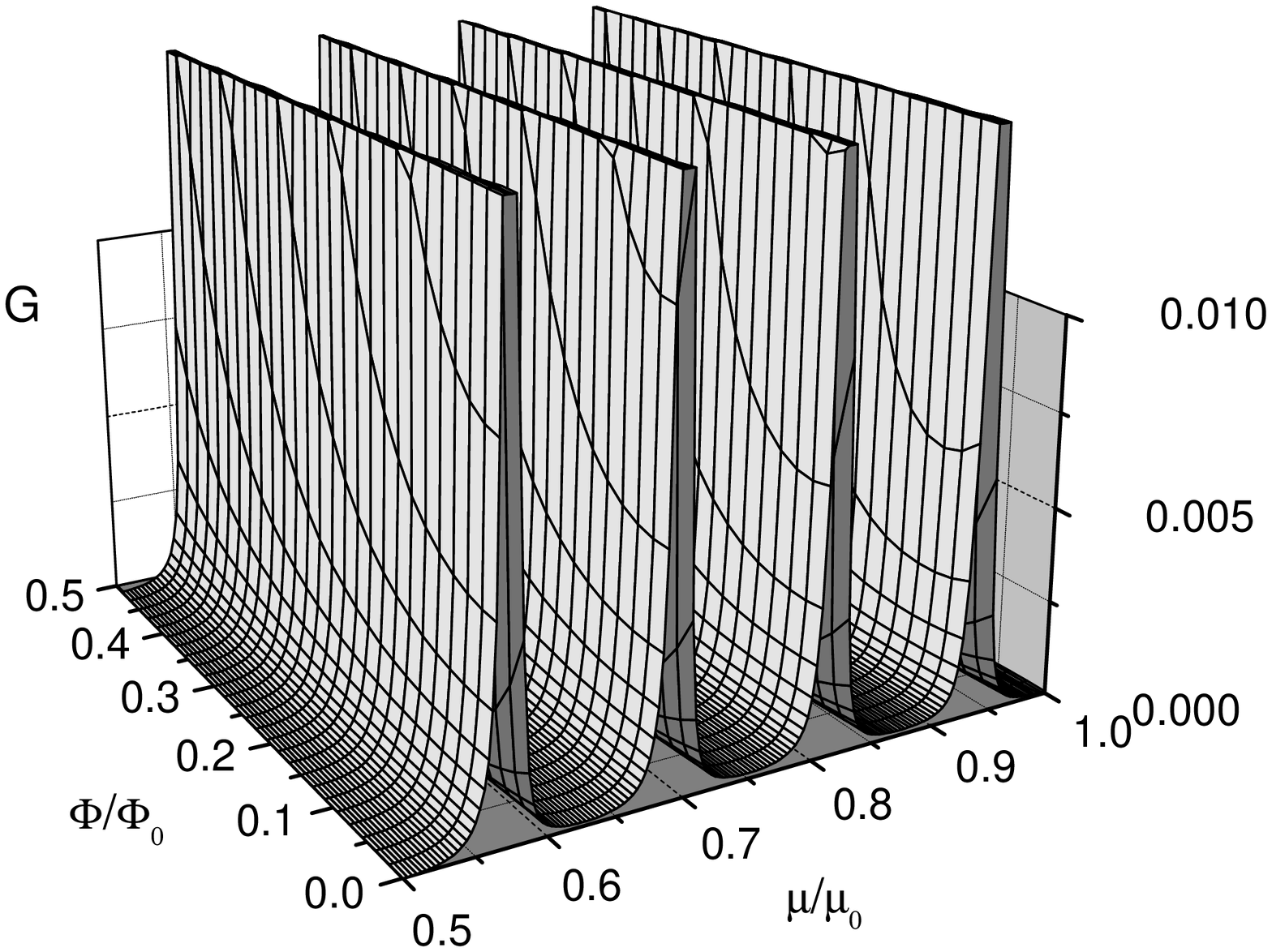}
\end{center}
\caption{Tunneling conductance $G$ of toroidal SWNT with $x=L/2$ (arbitrary
units).}
\label{fig1}
\end{figure}
The averages (\ref{eq:GMats}) are taken over the low-energy phase
Hamiltonian of toroidal SWNT \cite{Odintsov}, 
\begin{eqnarray}
H_{NT} &=&\int_{-L/2}^{L/2}\frac{dx}{2\pi }\left\{ \frac{v_{\nu \delta }}{%
K_{\nu \delta }}\left( \nabla \theta _{\nu \delta }-\frac{2K_{\nu \delta }}{%
v_{\nu \delta }}\mu \delta _{\delta +}\delta _{\nu +}\right) ^{2}\right. 
\nonumber \\
&&\left. +v_{\nu \delta }K_{\nu \delta }\left( \nabla \phi _{\nu \delta }-%
\frac{2e\Phi }{cL}\delta _{\delta +}\delta _{\nu +}\right) ^{2}\right\} .
\label{Hnt}
\end{eqnarray}
Here the standard interaction parameters $K_{\nu \delta }$ and velocities $%
v_{\nu \delta }$ of charge/spin ($\nu =\rho, \sigma$) excitations in
symmetric/antisymmetric ($\delta =\pm$) modes are introduced. We take into
account only the strongest interaction in charged mode, $K_{\rho +}\approx
0.2$, so that $v_{\rho +}=v_{F}/K_{\rho +}$ and $K_{\nu \delta }=1$, $v_{\nu
\delta }=v_{F}$ for three neutral modes. The electro-chemical potential $\mu$
(or gate voltage) is coupled to the charge density $\rho \propto \nabla
\theta _{\rho +} $ in SWNT. The magnetic flux $\Phi$ is coupled to the
current $I \propto \nabla \phi_{\rho +} $.

In order to describe the system with finite number of electrons, the fields $%
\theta _{\nu \delta }$ and $\phi _{\nu \delta }$ are decomposed into bosonic
and topological parts. The averages (\ref{eq:GMats}) over the corresponding
parts of the Hamiltonian (\ref{Hnt}) can be evaluated separately.

The conductance of the system displays resonant peaks as a function of the
electro-chemical potential $\mu $ (Fig. 1). The map of the peak positions in 
$\mu -\Phi $ plane depends on the wrapping vector and the length of toroidal
SWNT \cite{Odintsov}. The map for armchair SWNT of length $L=3na$ is shown
in Fig. 2 (here $a=2.46$ \AA\ is a lattice constant of graphite). The
diagram is periodic in $\mu $ with period $\mu _{0}=2\pi v_{\rho +}/K_{\rho
+}L$ and in $\Phi $ with the period $\Phi _{0}=h/e$. Eight electrons enters
SWNT per period $\mu _{0}$. The lines with the negative/positive slope
correspond to entering of electrons with orbital magnetic moment
parallel/antiparallel to the magnetic field. The slope of the lines $d\Phi
/d\mu =(\Phi _{0}/\mu _{0})/K_{\rho +}^{2}$ allows one to determine the
interaction in the charged mode \cite{Kinaret}. 
\begin{figure}[tb]
\par
\begin{center}
\leavevmode
\includegraphics[width=0.65\linewidth]{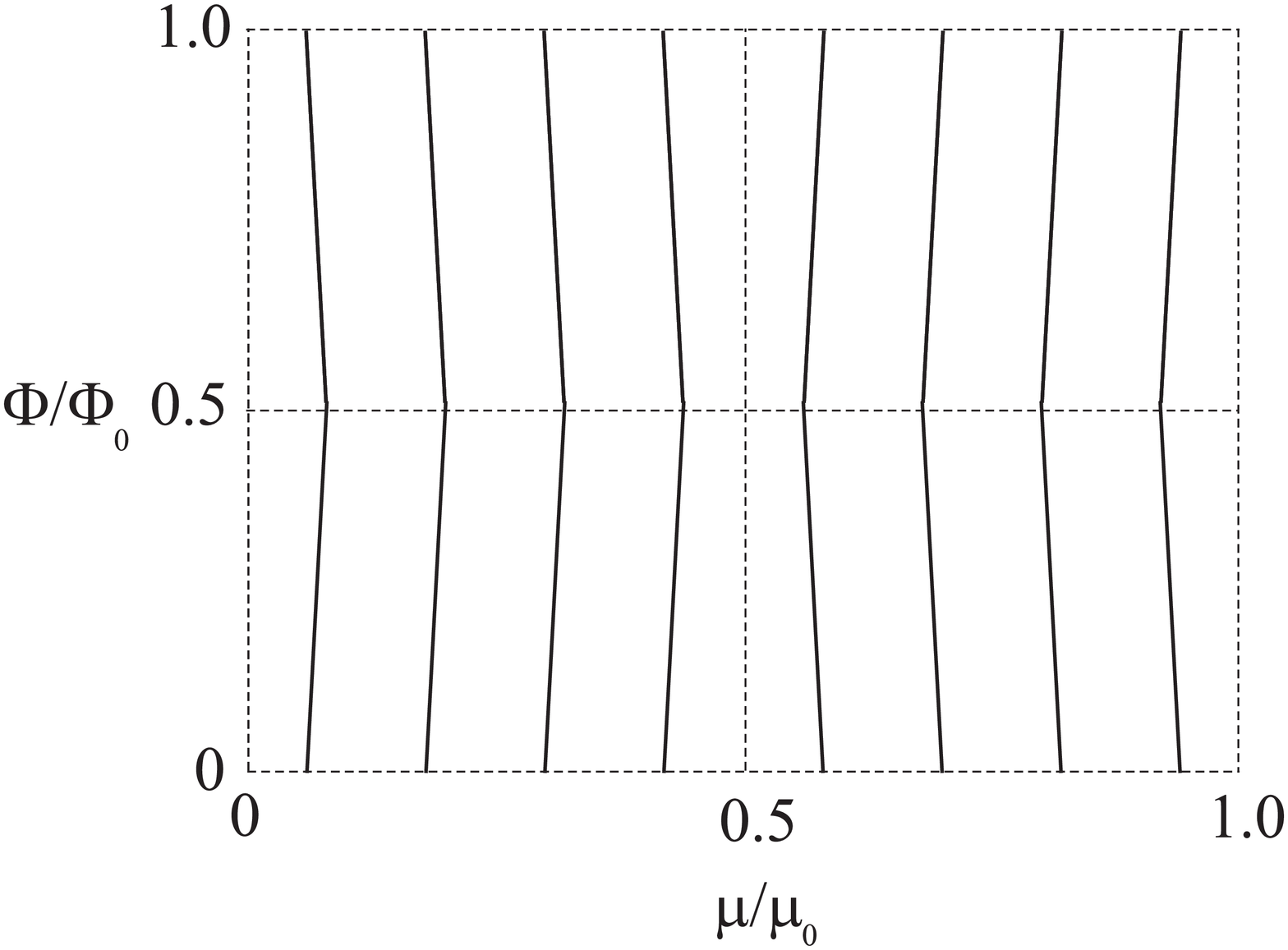}
\end{center}
\caption{The ground state of toroidal armchair SWNT of length $L=3na$}
\label{fig2}
\end{figure}

Close to the resonances, $|\Delta \mu |\ll \mu ^{\ast }$, the conductance
scales as $G\propto (a/L)^{2g}/\Delta \mu ^{2},$ $g=\sum_{\nu \delta
}(K_{\nu \delta }+1/K_{\nu \delta })/8$ (cf. Ref. \cite{Kinaret}). The
divergence at $\Delta \mu \rightarrow 0$ is an artifact of the perturbation
theory. At $x\ll L,$ in the energy range $\mu ^{\ast }\ll |\Delta \mu |\ll
\mu _{0}$ the conductance levels off as a function of $\Delta \mu $ and
shows power-law dependence on $x,$ $G\propto (a/x)^{2g-2}(a/v_{\rho +})^{2}.$
By comparing the two asymptotics of $G$ we obtain, $\mu ^{\ast }\simeq (2\pi
v_{\rho +}/L)(2\pi x/L)^{g-1}$. In addition, away from half-filling the
conductance shows oscillations as a function of $x$ with the period $\pi /q$
related to the mismatch $q\simeq 2\pi \mu /L\mu _{0}$ of the Fermi vectors
of right and left-moving electrons. At finite voltages and/or temperatures
we expect the competition of the considered process of coherent electron
transfer through SWNT with incoherent sequential tunneling through the
junctions.

We thank G.E.W. Bauer, C. Dekker, and U. Weiss for stimulating discussions.
The financial support of KNAW and kind hospitality of the University of
Stuttgart are gratefully acknowledged.


\begin{thebibliography}{9}
\bibitem{Liu}  J. Liu, et. al., Nature \textbf{385}, 780 (1997).

\bibitem{Avouris}  R. Martel, et. al., Nature \textbf{398}, 299 (1999).

\bibitem{Bockrath}  M. Bockrath, et.al. Nature \textbf{397}, 598 (1999).

\bibitem{Odintsov}  A.A. Odintsov, et. al., Europhys. Lett. \textbf{45},
598-604 (1999).

\bibitem{Kinaret}  J.M. Kinaret, et. al., Phys. Rev. B \textbf{57}, 3777
(1998).
\end{thebibliography}
\end{document}